# Effects of minor alloying on the mechanical properties of Al based metallic glasses

V. Jambur[a], C. Tangpatjaroen[a], J. Xi[a], J. Tarnsangpradit[a], M. Gao[a], H. Sheng[b] J. H. Perepezko[a], I. Szlufarska[a, *]

[a]*Department of Materials Science and Engineering, University of Wisconsin – Madison, WI 53706, USA*

[b]*Department of Physics and Astronomy, George Mason University, Fairfax, VA 22030, USA*

*Corresponding author:

Prof. Izabela Szlufarska

Department of Materials Science & Engineering

University of Wisconsin – Madison

Madison, WI 53706, USA

Phone: +1 (608) 265-5878

E-mail: szlufarska@wisc.edu

**Abstract**

Minor alloying is widely used to control mechanical properties of metallic glasses (MGs). The present understanding of how a small amount of alloying element changes strength is that the additions lead to more efficient packing of atoms and increased local topological order, which then increases the barrier for shear transformations and the resistance to plastic deformation. Here, we discover that minor alloying can improve the strength of MGs by increasing the chemical bond strength alone and show that this strengthening is distinct from changes in topological order. The results were obtained using Al-Sm based MGs minor alloyed with transition metals (TMs). The addition of TMs led to an increase in the hardness of the MGs which, however, could not be explained based on changes in the topological ordering in the structure. Instead we found that it was the strong bonding between TM and Al atoms which led to a higher resistance to shear transformation that resulted in higher strength and hardness, while the topology around the TM atoms had no influence on their mechanical response. This finding demonstrates that the effects of topology and chemistry on mechanical properties of MGs are independent of each other and that they should be understood as separate, sometimes competing mechanisms of strengthening. This understanding lays a foundation for design of MGs with improved mechanical properties.

1. Introduction

The unique combination of high strength and elasticity of metallic glasses (MGs) has attracted a lot of attention since the discovery of these materials[1,2]. Several MGs with a wide range of compositions have been synthesized to date[3] and minor alloying has emerged as common way to manipulate their mechanical properties[4–13]. It is therefore important to understand how small amounts of alloying elements can have such a strong influence on the mechanical behavior of MGs.

Unlike crystalline metals, the mechanical response of MGs is not controlled by discernible structural features such as dislocations, stacking faults, and grain boundaries. The carriers of deformation in metallic glasses are called shear transformation zones (STZ)[14]. These are local clusters of atoms that undergo an inelastic shear distortion to accommodate plastic strains during deformation. STZs percolate along the planes of the maximum shear stress in the sample and eventually form regions of localized deformation called shear bands[2,15]. The activation of STZs and formation of shear bands is influenced by the atomic level structure in the MGs.

Even though MGs lack long-range structural order, they possess short- and medium-range orders, which can influence mechanical properties. Shi and Falk[16] studied the mechanical response of a binary metallic glass using the Lennard-Jones potential and proposed that metallic glasses derived their strength from a backbone of atoms with local short-range order (SRO). This report was followed by several other studies that investigated the influence of the atomic structure on the mechanical behavior of a more realistic system such as Cu-Zr and confirmed that indeed these MGs derived their mechanical strength from local SRO[6,17,26,18–25]. Atoms arranged in icosahedral clusters were shown to be more resistant to shear transformations than other groups of atoms with geometrically unfavored motifs (GUMs). The GUMs are disordered regions with liquid like structure and act as soft spots with a lower resistance to shear transformations and fertile sites for activation of STZs[27,28]. A higher fraction of icosahedral clusters in the structure was therefore associated with a higher resistance to plastic deformation.

The effects of chemical bonding on the mechanical behavior of metallic glasses have been studied in the context of MGs with significant fractions (~20% or more) of metalloids[29–31]. For example, in Pd-Si MGs, the covalent bonds associated with Si atoms were found to be more difficult to break than metallic bonds associated with Pd atoms. Since the fraction of Si atoms is significant in these alloys, Si atoms can suppress cavitation in Pd-Si, promoting a larger fracture toughness[29].

The effects of minor (i.e., on the order of a few percent) alloying elements on mechanical properties of MGs have been typically explained in terms of the effect that minor elements have on the atomic configurations and the topological order[5–9,11,12]. For instance, the addition of small amounts of Al to Cu-Zr MGs has been shown to increase the resistance to plastic flow due to the strong, covalent-like bonding between Cu and Al atoms, which in turn leads to shortening of Al-Cu bond lengths and an increase in the icosahedral order[6]. In other words, the effect of the chemical bonding has been conflated with the increased topological order.

The goal of this work is to understand the role of minor alloying in the strengthening of Al rich MGs. Our interest in these MGs is motivated by their light weight, high specific strength[32,33], good corrosion resistance[34], and thermoplastic forming ability[35], which makes them excellent candidates for applications in such areas as micro- and nano-scale devices[35], wear and corrosion resistant coatings[36], composites with excellent mechanical and tribological properties[37]. We have chosen Al-Sm based glasses alloyed with different transition metals (TM = Cu, Ag, Au) for this study. The Al-Sm based glasses serve as a good model system since they are one of the better understood Al-based glasses[35,38–41] and have been previously shown to possess good GFA and good stability due to the icosahedral SRO[40]. We have found that minor alloying of Al-Sm with TMs increases the strength of the MG, but this effect is due to the change in the chemical bond strength alone and cannot be explained by changes in topological order. In fact, we found that the effects of topology and the chemical bond strength are independent of each other in controlling the mechanical response of MGs subjected to minor alloying. Mechanical properties of the Al-Sm based glasses have been investigated experimentally using nanoindentation. The experiments are complemented by a detailed analysis of the atomic level structure using classical as well as *ab initio* molecular dynamics (MD).

## 2. Methods

### 2.1 Simulations

Classical molecular dynamics (MD) simulations were carried out with LAMMPS software package[42]. Embedded Atom Method (EAM) potential for Al-Sm-Cu ternary system was developed specifically for this study. The parameters for the potential are summarized in the supplementary material. Amorphous structures of $Al_{92}Sm_8$ and $Al_{90}Sm_8Cu_2$ were obtained by starting out with an FCC crystal lattice containing ~11000 atoms with periodic boundary conditions in all three directions. The atoms were assigned random initial velocities from a Gaussian distribution and equilibrated in the constant energy – constant volume (NVE) ensemble at a temperature of 300K for 20ps. The system was then equilibrated in the constant pressure – constant temperature (NPT) ensemble at 300K for another 200ps. This was followed by heating to a temperature of 2000K in the NPT ensemble followed by equilibration for 2ns. The melted sample was then quenched down to a temperature of 50K in the NPT ensemble. The quenching was carried out in steps of 50K with a cooling rate of $10^{10}$K/s. The melt quenched samples with dimensions of ~ 6nm x 6nm x 6nm were replicated to obtain larger samples with dimensions ~ 120nm x 120nm x 6nm, containing ~4.5 million atoms. The replicated samples were then equilibrated at 50K for 100ps. This was followed by uniaxial tensile deformation at a constant strain rate of $2 \times 10^7$/s at 50K. OVITO[43] software was used for Voronoi analysis and, calculation of atomic strains and non-affine displacements in the classical MD samples.

*Ab initio* molecular dynamics (AIMD) simulations were performed in the framework of the density functional theory (DFT) using the Vienna Ab-Initio Simulation Package (VASP)[44]. Projector-augmented-wave (PAW) potentials[45] were used to mimic the ionic cores, while the generalized gradient approximation (GGA) in the Perdew-Burke-Ernzerhof (PBE)[46] approach was employed for the exchange and correlation functional. The integration over the Brillouin zone was performed using the $\Gamma$ point for AIMD simulations. To prepare a model of glassy alloys for the AIMD simulations, we first created a glassy $Al_{92}Sm_8$ system with 256 atoms using classical MD simulations; the settings of the classical MD simulations are discussed elsewhere[47]. After that, we substituted Al with X=Ag, Au, and Cu by randomly replacing atoms in the liquid phase to obtain the exact composition, up to 2% of the minor alloying elements. The resulting configurations were further equilibrated in AIMD at 1300 K for 3 ps in the constant volume – constant temperature (NVT) ensemble with the Nose-Hoover thermostat and a time step of 3 fs. Afterwards, the systems were subsequently cooled down to 300 K. The equilibrium volume at particular temperature was obtained by constructing pressure-volume (*P-V*) equation of state[48]. After determining the volume, the samples were equilibrated for 15 ps at 300 K with their equilibrium volumes. 20 melt quenched samples with different initial atomic configurations were prepared for each sample to obtain statistically relevant results. In order to improve statistics, in these isothermal simulations, the first 12 ps of the simulation had been treated as the equilibration period run, whereas the remaining steps were treated as the production run to calculate the average values of icosahedral fraction using the Voronoi analysis tool in OVITO[43].

We used DFT calculations to determine the force constant for one specific atom with its surrounding environment. Based on the harmonic approximation, the force constant can be described through the equation of $k = m_X f^2$, here $k$ is the force constant, $m_X$ is the atomic mass of the doping atom (X=Cu, Au, and Ag), and $f$ is the vibrational frequency. Vacancy formation energy for atom X is calculated through the equation of $E_f^X = E_{defect} - E_{perfect} + \mu_{atom,X}$. Here, $E_f^X$ is the energy cost for the formation of one isolated X atom, and $E_{defect}$ and $E_{perfect}$ are the system with and without X vacancy, respectively. $\mu_{atom,X}$ is the isolated atomic energy, which is calculated by inserting one X atom into a vacuum box.

**2.2 Experimental**

Four Al-Sm based metallic glass compositions were selected for this study - $Al_{92}Sm_8$, $Al_{90}Sm_8Cu_2$, $Al_{90}Sm_8Ag_2$, and $Al_{90}Sm_8Au_2$. All precursors were prepared using arc melter method in an argon atmosphere. The pure elements for each composition were melted and stirred together at least five times to ensure homogeneity. The chemical composition of the ingots was confirmed by Energy Dispersive Analysis (EDS, scanning electron microscope Zeiss LEO 1530). Following this, the precursors were subjected to melt spinning on a copper wheel at the wheel speed of 55m/s, in an argon atmosphere to produce MG ribbons. The chemical composition of the ribbons was also confirmed by EDS. The amorphous structure of the as-spun ribbons was ascertained by X-ray diffraction (Bruker D8 Discover Diffraction with Cu Kα radiation).

For nanomechanical testing, the MG ribbons were placed in a custom-made aluminum mount and polished mechanically using abrasive SiC grit paper followed by 1 µm diamond suspension. The hardness of the samples was evaluated using the Hysitron TI 950 TriboIndenter equipped with a diamond Berkovich tip probe. A series of indents (120 – 130 for each composition) were performed, with a 200 µN/s loading rate, a 2 s hold at the maximum load, and a 200 µN/s unloading rate, with the maximum loads ranging from 900 to 2400 µN. Continuous stiffness measurement using Bruker nanoDMA III transducer in CMX mode was done to measure the hardness as a function of the indentation depth. The measurement was done at a constant strain rate and a frequency of 220Hz. The sample was loaded to a peak load of 10mN.

3. **Results and discussion**

**3.1 Effect of icosahedral order on the mechanical behavior of binary Al-Sm metallic glasses**

Before looking into the effects of minor alloying on the mechanical properties of Al-Sm MGs, we determine what controls the mechanical behavior of binary Al-Sm glasses using classical MD simulations. Melt quenched samples with compositions $Al_{90}Sm_{10}$, $Al_{92}Sm_{8}$, and $Al_{94}Sm_{6}$ were prepared using the procedure described in section 2.1 and the structure was studied using Voronoi analysis. In this technique, each atom is considered as a center of a Voronoi polyhedron (VP) and assigned Voronoi indices that describe shapes of the VP faces. In all three Al-Sm samples considered in our study, the majority of VPs that are centered on Al atoms were found to have indices <0,0,12,0> and <0,1,10,2> (Supplementary Fig.1). While VPs with indices <0,0,12,0> represent full icosahedral SRO, <0,1,10,2> can be considered distorted icosahedral SRO. Distorted icosahedra refer to VPs with a large number of fivefold bonds with coordination numbers close to 12, such as <0,1,10,2>, <0,2,8,2>, and <0,2,8,1>[49]. We calculated the fraction of VPs with more than 8 pentagonal faces and coordination numbers in the range 11-13 as a measure of the degree of icosahedral order and found that it increases monotonically with the atomic fraction of Sm in the alloys.

The three alloys were then subjected to uniaxial tensile deformation to obtain tensile stress – strain curves (Fig.1a). The peak stress in these curves corresponds to the resistance to the initiation of plastic flow. We can see that the peak stress increases with the increasing degree of icosahedral order (Fig.1b). After the stress reaches its peak value, it drops to a relatively constant flow stress in all the samples. The flow stress corresponds to the resistance to plastic flow offered by the rejuvenated/disordered structure in the shear bands. The difference between the peak stress and the flow stress represents the difference in the strengths of the initial structure with icosahedral order and the rejuvenated structure in the shear bands. This difference is an indicator of the propensity for localization of deformation[6]. As shown in Fig.1a, the drop in the stress value is the largest in the case of $Al_{90}Sm_{10}$, followed by $Al_{92}Sm_8$. The change from the peak stress to the flow stress in $Al_{94}Sm_6$ seems more gradual. This indicates that $Al_{90}Sm_{10}$ has the largest propensity for localization of deformation, followed by $Al_{92}Sm_8$, and then $Al_{94}Sm_6$. The difference in the degree of localized deformation in $Al_{90}Sm_{10}$ and $Al_{94}Sm_6$ can be seen in Fig.2. These results show that a higher degree of icosahedral order leads to a higher resistance to plastic flow and a higher propensity for localized deformation in binary Al-Sm alloys, consistent with what has been previously reported in Cu-Zr alloys[6,21].

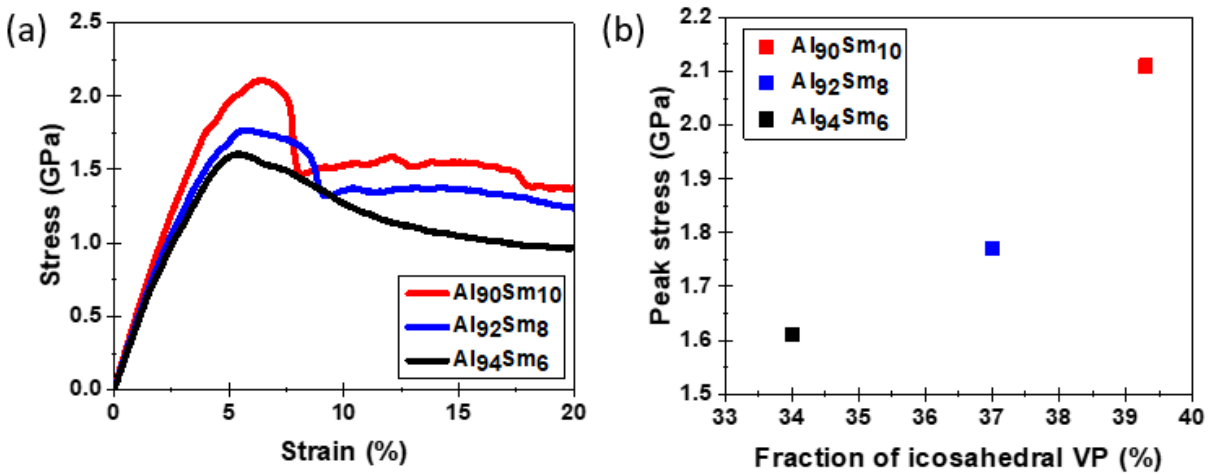

*Fig.1: Mechanical response of binary Al-Sm MGs. (a) Simulated tensile stress strain curves for binary Al-Sm MGs at 50K. The change from the gradual drop in stress after the peak stress in $Al_{94}Sm_6$ to the steep drop in $Al_{90}Sm_{10}$ indicates the increasing propensity for localized deformation. (b) Correlation between peak stress and the fraction of icosahedral VP in the binary Al-Sm alloys.*

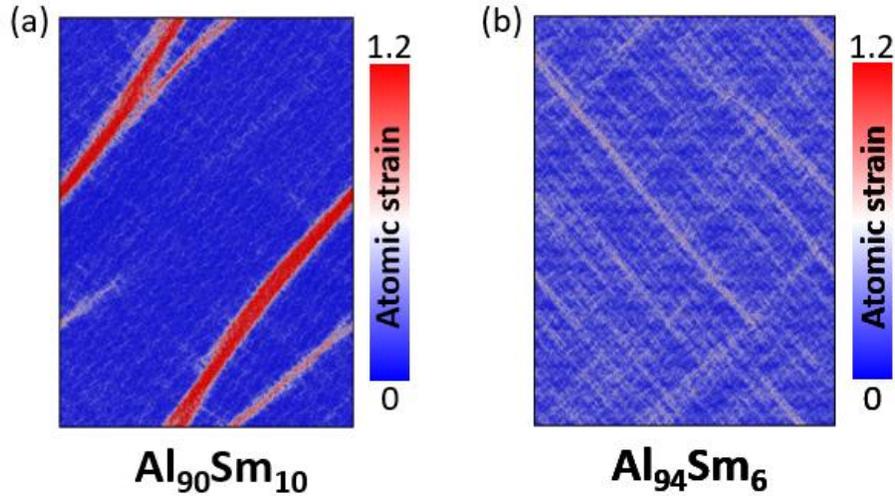

*Fig.2*: ***Atomic strains during tensile deformations.*** *Deformation is more localized in **(a)**$Al_{90}Sm_{10}$ compared to **(b)**$Al_{94}Sm_{6}$.*

**3.2 Effect of TM addition on mechanical properties**

The effects of minor alloying with TM on mechanical properties have been first studied experimentally. Three different TMs (Cu, Ag, Au) were substituted in place of Al in $Al_{92}Sm_8$ to obtain alloys with compositions $Al_{90}Sm_8Cu_2$, $Al_{90}Sm_8Ag_2$, and $Al_{90}Sm_8Au_2$. Experimental ribbon samples with the above compositions were prepared by melt spinning and the presence of amorphous structure was confirmed using x-ray diffraction (Supplementary Fig.2). Nanoindentation experiments were performed on the melt – spun ribbon samples to determine hardness and elastic modulus using the Oliver – Pharr method[50] (Fig.3a). The hardness of the ternary compositions with minor amounts of TMs was significantly higher than that of the binary alloy. However, the elastic modulus appeared to be lowered after minor alloying. Measurement of the elastic modulus as a function of the indentation depth using the continuous stiffness method (CSM)[51] revealed that the elastic moduli of the ternary compositions were initially slightly higher than that of the binary alloy. However, the modulus decreases rapidly with an increasing indentation depth and eventually crosses over with the elastic modulus of the binary alloy (Fig.3b). This means that the ternary alloys experience more strain-induced softening than the binary alloy. The same behavior was also seen in the hardness measured using the CSM method, but the initial difference in hardness between the ternary alloys and the binary was too large for them to cross over due to softening (Supplementary Fig.3). The large hardness in the ternary alloys is indicative of a higher resistance to plastic flow, whereas the more pronounced softening means that there is more localized deformation. The effect of minor alloying appears similar to the effect of increasing the degree of icosahedral order in the binary Al-Sm alloys and our first hypothesis is that the TMs increase the icosahedral order in the Al-Sm alloys. This hypothesis can be tested using atomistic simulations.

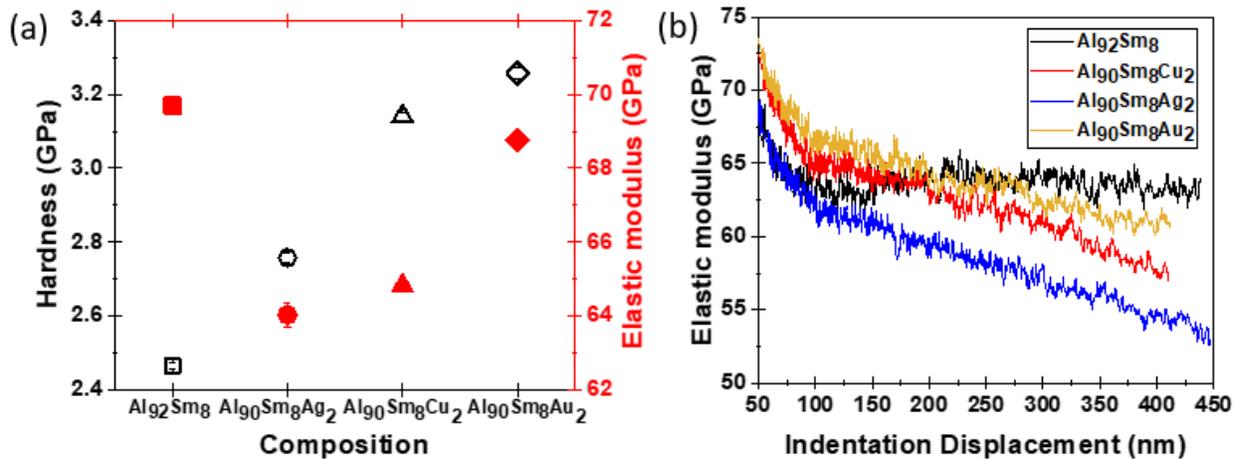

*Fig.3: Effect of TM addition on mechanical response of Al-Sm MGs. (a) Experimentally measured nanoindentation hardness (black, open symbols) and elastic modulus (red, closed symbols) for $Al_{92}Sm_8$, $Al_{90}Sm_8Ag_2$, $Al_{90}Sm_8Cu_2$, and $Al_{90}Sm_8Au_2$ ribbon samples. Hardness increases after minor alloying with TMs. Error bars represent standard errors. (b) Elastic modulus as a function of the indentation depth measured using the continuous stiffness method (CSM). There is a crossover between the moduli of the binary and the ternary alloys as a function of depth due to a more pronounced softening in the ternary alloys.*

## 3.3 Effect of TM addition on atomic level structure

To investigate the effect of the TMs on the atomic-level structure of Al-Sm MGs, we prepared melt-quenched structures of $Al_{92}Sm_8$, $Al_{90}Sm_8Ag_2$, $Al_{90}Sm_8Cu_2$ and $Al_{90}Sm_8Au_2$ alloys using *ab initio* MD simulations based on the density functional theory (DFT). After quenching to 300K, we performed Voronoi analysis to find the most common VP in the four alloys (Fig.4a). We observed that minor alloying did not have a large effect on the atomic level structure of the MGs. We then calculated the degree of icosahedral ordering in the four alloys using the same criteria as we have used earlier to analyze binary Al-Sm alloys prepared in classical MD simulations. We found that the increase in hardness with minor alloying could not be explained based on the changes in the amount of icosahedral ordering in the structure (Fig.4b). In addition, the local topological order around the different TM atoms was found to vary, depending on the size of the TM atoms but this trend still does not explain the measured changes in mechanical properties of the MGs (Supplementary Fig.4). To understand the effect of minor alloying, it is therefore necessary to analyze the mechanical behavior of these MGs at the atomic level.

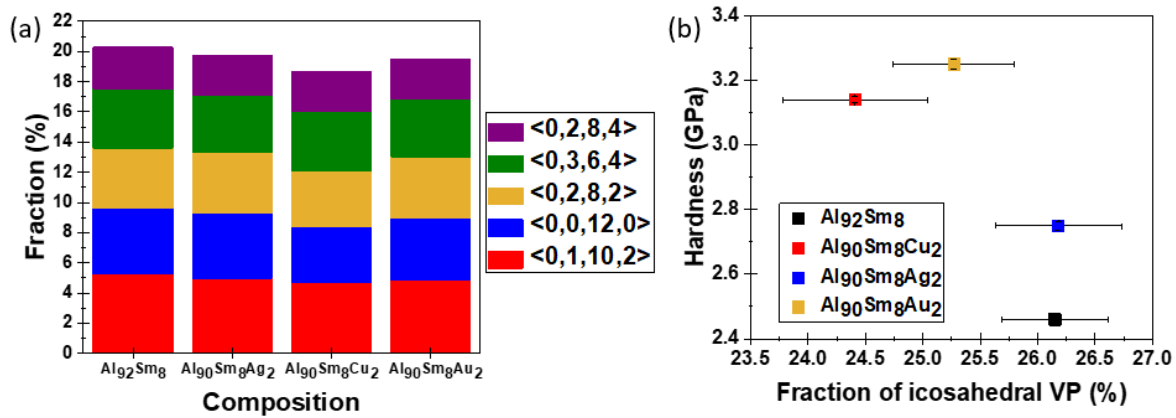

***Fig.4***: *Effect of TM addition on topological short-range order in Al-Sm MGs. (**a**) Most common VP in the Al-Sm-TM MGs at 300K obtained from ab initio MD simulations. (**b**) Experimental nanoindentation hardness plotted against the fraction of icosahedral VP at 300K (Error bars*

*represent standard error). The higher hardness in the ternary alloys despite having fewer icosahedral VP means that the correlation between the degree of icosahedral ordering and mechanical properties does not hold in the ternary alloys.*

**3.4 Mechanical behavior at the atomic level**

Classical MD simulations of tensile deformation were carried out on melt-quenched $Al_{90}Sm_8Cu_2$ to study the effect of TM atoms on the mechanical response of the Al-Sm-TM MGs. The tensile stress – strain curves of the binary $Al_{92}Sm_8$ and $Al_{90}Sm_8Cu_2$ are plotted in Fig.5a and they show that $Al_{90}Sm_8Cu_2$ has a higher peak stress and a higher drop in stress from to peak stress to the flow stress than the binary. This means that the ternary has more resistance to plastic flow and more localized deformation, which is consistent with the conclusion drawn from the nanoindentation experiments. This agreement between MD simulations and experiments further validates the use of our EAM potential for analysis of the role of TM in deformation of Al-Sm-TM glasses. Furthermore, we have performed the Voronoi analysis on samples generated from classical MD simulations and found that while the icosahedral order was still dominant in the ternary $Al_{90}Sm_8Cu_2$ alloy, the degree of icosahedral order was lower than that in the binary $Al_{92}Sm_8$ alloy. This trend is consistent with the results from *ab initio* calculations shown in Fig.3b.

At the atomic scale, a higher resistance to plastic flow can be attributed to a higher resistance to the activation of STZs, which are the carriers of plastic deformation. In order to determine the reason for a higher resistance to the activation of STZs in the $Al_{90}Sm_8Cu_2$ system, we used the non–affine squared displacement ($D^2_{min}$)[14] parameter to identify those atoms that underwent local irreversible shear transformations to form STZs. We have subsequently identified the 5% of atoms in the sample that have the highest and the lowest $D^2_{min}$ at various stages during tensile deformation. We found that the atoms with the highest $D^2_{min}$ were all concentrated along a single plane that constituted the shear band (Fig.5a inset), whereas the atoms with the lowest $D^2_{min}$ were distributed uniformly outside the shear bands. Among the 5% of atoms with the highest $D^2_{min}$, the fraction of Cu atoms was found to be lower than the overall fraction of 2%. At the same time, among the 5% of atoms with the lowest $D^2_{min}$, the fraction of Cu atoms was found to be significantly higher than the overall fraction of 2% (Fig.5b). These results indicate that Cu atoms have a lower propensity to undergo local shear transformations as compared to Al and Sm atoms. Therefore, clusters of atoms containing Cu offer more resistance to the activation and percolation of STZs to form shear bands, which in turn leads to higher strength and hardness.

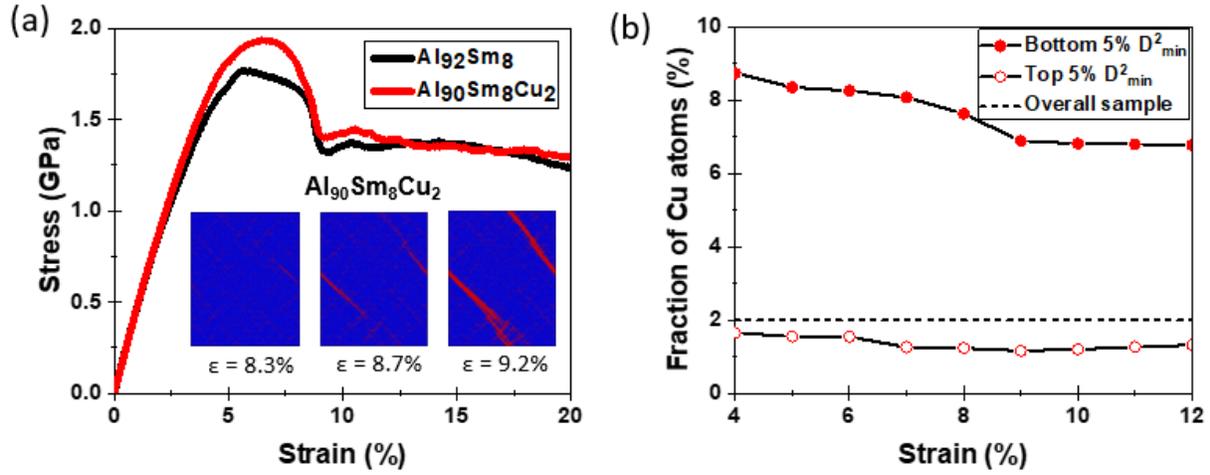

*Fig.5: Role of Cu in mechanical behavior of Al-Sm MGs. (a) Tensile stress – strain curves for $Al_{92}Sm_8$ and $Al_{90}Sm_8Cu_2$ from MD simulation at 50K; Inset: Non affine squared displacement ($D^2_{min}$) map showing formation of shear band in $Al_{90}Sm_8Cu_2$ (Red atoms are the 5% of atoms with the highest $D^2_{min}$ values). (b) Fraction of Cu among the 5% of atoms with the lowest and highest $D^2_{min}$ values compared with the overall fraction of Cu.*

To answer the question of why Cu atoms provided more resistance to shear transformations, we first considered the influence of topological order around the Cu atoms. The most common types of VPs around Cu atoms are shown in Fig.6a. Unlike Al atoms, the smaller Cu atoms are dominated by distorted icosahedral VP with indices <0,2,8,1>, followed by VPs with indices <0,2,8,0>, which are sometimes referred to as bi-capped square Archimedean antiprism (BSAP)[49]. The two most common VPs centered on Cu atoms have coordination numbers of 11 and 10, respectively. The distorted icosahedral VP with indices <0,2,8,1> and BSAP with indices <0,2,8,0> are known to be the most stable and efficiently packed VP for their respective coordination numbers[49]. We next looked at the mechanical response of these two types of VPs to see if they had any influence on the propensity of Cu atoms to undergo local shear transformations. For this purpose, we compared the fractions of these VPs among Cu atoms that were part of the bottom 5% of $D^2_{min}$ values with the overall fraction of these VP among all the Cu atoms in the samples and we found these fractions to be almost the same (Fig.6b). In contrast, a similar comparison for dominant Al centered VPs (with indices <0,0,12,0> and <0,1,10,2>) showed that their fraction among the bottom 5% of $D^2_{min}$ values was significantly higher than the overall fraction of these VPs in the sample. This means that while the Al atoms with icosahedral SRO were more resistant to shear transformations than other Al atoms (Fig.6c), the SRO does not play the same role in the case of Cu atoms. In other words, the topological order does not have any significant influence on the resistance to deformation provided by Cu atoms in $Al_{90}Sm_8Cu_2$, but the topological order plays a key role in determining the resistance to deformation provided by Al atoms in this MG.

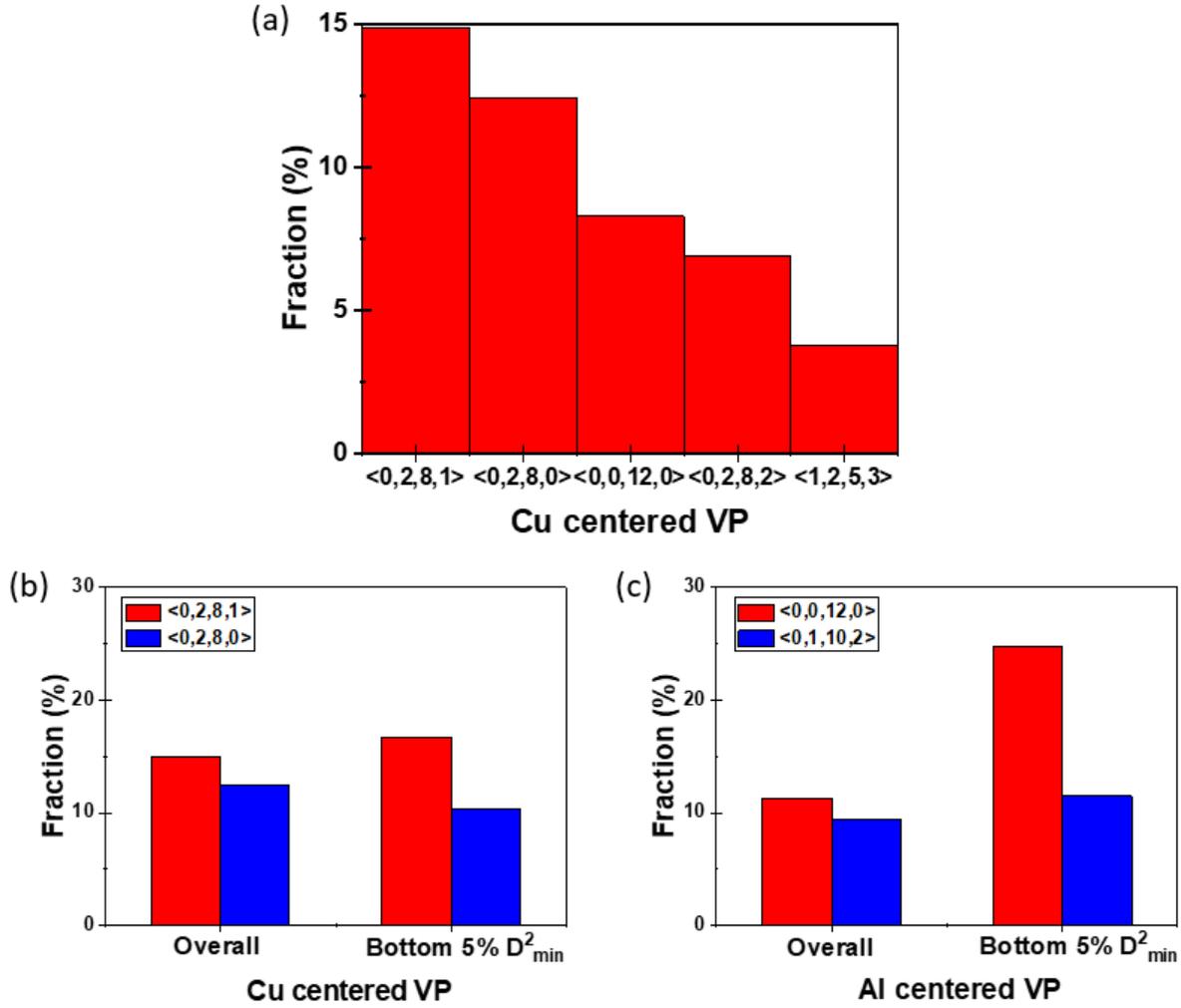

*Fig.6: Effect of topological order in mechanical response of Cu and Al atoms. (a) Most common VP around Cu atoms in $Al_{90}Sm_8Cu_2$. Fraction on the vertical axis indicates the fraction of each type of VP among all the Cu atoms in the system. The overall fractions of the dominant clusters around (b) Cu atoms and (c) Al atoms are compared with their fractions among the 5% of atoms with the lowest $D^2_{min}$ (as determined by classical MD). Overall fractions are normalized to the total number of each type of atom in the entire sample while fractions in the bottom 5% of $D^2_{min}$ are normalized to the number of each type of atom in the bottom 5% of $D^2_{min}$.*

### 3.5 Correlation between Al-TM bond strength and mechanical properties

So far, we have demonstrated that Cu atoms provide a higher resistance to local shear transformations as compared to Al atoms and that the resistance of Cu atoms to deformation is independent of their topological environment. This finding suggests that the nature of bonding between Cu atoms and the surrounding Al atoms could be the main reason for their higher resistance to local shear transformations. We hypothesize that the mechanical strength of Al-Sm-TM alloys is correlated with the strength of the Al-TM bond and that this correlation is strong enough to break the correlation between mechanical response and the dominant topological SRO in the MG.

To test this hypothesis, we calculated the force constant of the atomic vibrations of TM atoms as well as the formation energy of a TM vacancy in the melt quenched samples, as two measures of the Al-TM bond strength. Here, vacancy formation energy is defined as the energy cost of isolating a TM atom by breaking the bonds between the atom and its neighbors. These calculations were carried out using *ab initio* calculations on amorphous samples that were also prepared using *ab initio* methods. In Figs.7a and 7b we plot the experimentally measured hardness against the calculated force constant and the vacancy formation energy, respectively. We see that the hardness increases monotonically with the force constant and with the vacancy formation energy. The difference in the Al-TM bond strengths is also reflected in the relative shortening of the Al-TM bond lengths for the different TMs due to the partially covalent nature of the bond[52–55]. We have used pair distribution functions from the *ab initio* MD simulations of $Al_{90}Sm_8Cu_2$, $Al_{90}Sm_8Ag_2$, and $Al_{90}Sm_8Au_2$ MGs to calculate the Al-TM bond lengths and compared them with the bond lengths calculated theoretically by assuming purely metallic bonding between the Al and TM atoms (as explained in Table 1). We found that Al-Au bonds had the largest percentage shortening in bond length (7.7%) and therefore the strongest bonds, followed by Al-Cu (6%), and then Al-Ag (4.2%), which is in agreement with the trends in the force constants and in the vacancy formations energies determined from *ab initio* calculations.

This analysis confirms that the mechanical properties of Al-Sm-TM alloys are correlated with the strength of the Al-TM bonds.

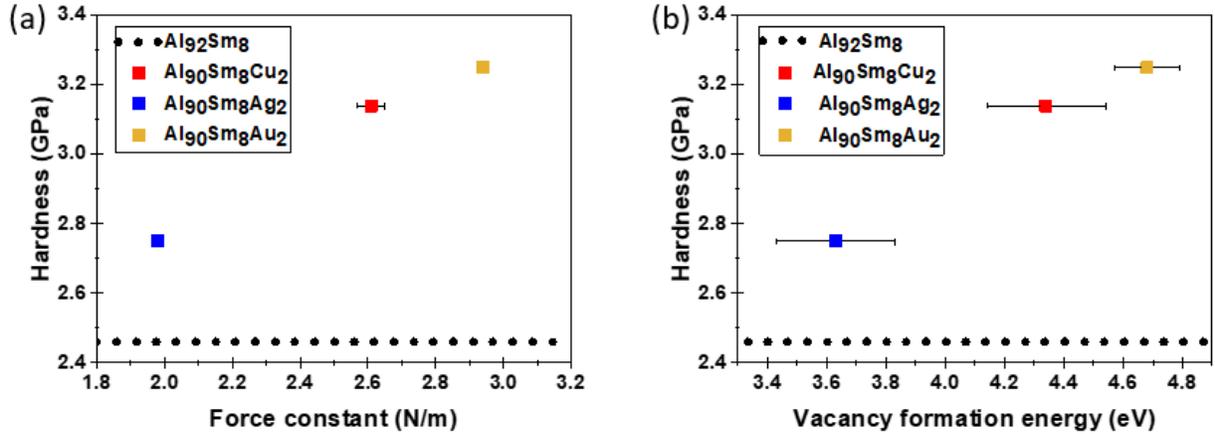

*Fig.7: Relation between hardness and Al-TM bond strength.* Trends between the experimentally measured hardness and *(a)* force constants of Al-TM bonds along x normal direction (the force constants of Al-Tm bonds along the other two normal directions have the similar trend) and *(b)* TM vacancy formation energies determined from ab initio simulations. The dotted line is added to show hardness of $Al_{92}Sm_8$. (Error bars represent standard errors).

| Al metallic radius[1] (Å) | TM | TM metallic radius[56] (Å) | Sum of metallic radii (Å) | Al-TM RDF peak (Å) | % shortening in bond length |
|---|---|---|---|---|---|
| 1.43 | Cu | 1.28 | 2.71 | 2.55 | 6.0 |
| 1.43 | Ag | 1.44 | 2.87 | 2.75 | 4.2 |
| 1.43 | Au | 1.44 | 2.87 | 2.65 | 7.7 |

*Table 1: Al-TM bond lengths calculated from ab initio PDF compared with the bond length obtained as sum of metallic radii.*

## 4. Conclusion

In conclusion, we have shown that the strengthening of Al-Sm MGs with minor alloying is due to strong chemical bonds between Al and TM atoms rather than due to changes in the atomic configurations and the topological order. Further, our results show that stronger chemical bonding between atoms does not necessarily result in a more ordered structure and that the effect of chemical bonding and the effect of topology may sometimes compete with each other in controlling the mechanical properties of the MG. This is in contrast to Cu-Zr MGs alloyed with Al where the effect of chemical bonding between Cu and Al is conflated with increased icosahedral order[6]. Our results therefore point to the importance of studying chemical and topological effects separately in MGs and also highlight the necessity of studying alloy systems other than the commonly studied Cu-Zr system to provide understanding of the structure-property relations in MGs.

**Acknowledgement**

This research was primarily supported by NSF through the University of Wisconsin Materials Research Science and Engineering Center (DMR-1720415). H.S. acknowledges support by the NSF under Grant No. DMR-1611064. M.G. and J.H.P. acknowledge the financial support from the Office of Naval Research (N00014-16-1-2401).

**Competing interests**

The authors declare no competing financial interests.